\documentclass{article}

\usepackage{arxiv}

\usepackage[utf8]{inputenc} % allow utf-8 input
\usepackage[T1]{fontenc}    % use 8-bit T1 fonts
\usepackage{hyperref}       % hyperlinks
\usepackage{url}            % simple URL typesetting
\usepackage{booktabs}       % professional-quality tables
\usepackage{amsfonts}       % blackboard math symbols
\usepackage{nicefrac}       % compact symbols for 1/2, etc.
\usepackage{microtype}      % microtypography
\usepackage{lipsum}
\usepackage{graphicx}
\usepackage{amsmath}
\usepackage{subcaption}
\graphicspath{ {./images/} }

\title{Analytic expressions for estimation of the critical properties of inhomogeneous Ising models}

\author{
 Vladislav Egorov \\
  Center of Optical Neural Technologies,\\
  Research Center “Kurchatov Institute” - Scientific Research Institute of System Analysis\\
  Moscow, Russian Federation \\
  \texttt{rvladegorov@rambler.ru} \\
   \And
 Stepan Osipov \\
  Laboratory of Mathematical and Computer Modelling of Nanostructures,\\
  Cherepovets State University\\
  Cherepovets, Russian Federation \\
  \texttt{alan.turing1912@yandex.ru} \\
}

\begin{document}

\maketitle

\begin{abstract}
In many applications of spin models, the fast estimation of their critical temperatures and other physical properties is of great importance. In this work, we present the analytical expressions estimating the critical properties of inhomogeneous Ising models with ferromagnetic interactions. The expressions were obtained within the framework of the m-vicinity method. The accuracy of the critical temperature estimations was evaluated through comparison with Monte Carlo simulations. Special attention was given to the case when the model consists of two interacting interpenetrating homogeneous sublattices, and relationships for the compositional dependence of the critical temperature were derived.
\end{abstract} %%%%%%%%%

\section{Introduction}\label{Intro}

The Ising model, which was originally formulated to describe the ferromagnetic phase transition, nowadays has countless applications in different scientific fields. The very wide spectrum of its applications includes solving NP-optimization problems \cite{lucas2014ising, takabatake2022solving}, econophysics \cite{valle2021equity, bury2013market, borysov2015us} and sociophysics \cite{mullick2025sociophysics}.

Particularly, we should mention applications of Ising models to studying neural networks \cite{amit1985spin, karandashev2012weighted}. These studies are based on the fact that deep neural networks can be mapped into the Ising model. In the work \cite{stosic2022ising}, it was shown that the critical temperature and the width of the density of states have higher values in better trained models. Therefore, the performance of the neural network can be evaluated by applying the methods of statistical thermodynamics. 
 
Classical applications of the Ising model to describe phase transitions in ferroics are also still relevant \cite{yuksel2023exploring, de2025simple}. Special interest is now given to inhomogeneous lattices with randomness \cite{li2024dynamic} and/or competing interactions \cite{lee2024frustrated, abalmasov2024local, rahmatullaev2023ground}. 
Also, active research is going on the critical behaviour of Ising-like systems consisting of two interacting interpenetrating sublattices \cite{elmghabar2022monte, elidrysy2025comparative}.

Many of the above mentioned applications require the accurate estimation of critical properties. The main problem is that an exact solution for the Ising model can be found only in a few particular cases, which mainly include 1d, 2d lattices \cite{baxter2016exactly} and models with infinite range interactions \cite{roberts2023exact}. Therefore, Monte Carlo (MC) methods such as the Metropolis algorithm \cite{haggkvist2004monte} and, more recently, the Wang-Landau algorithm \cite{wang2001efficient} remain the main tools to study three- and higher-dimensional spin models as well as inhomogeneous lattices. However, such a technique requires extensive computer simulations especially when dealing with large lattices and long range interactions. For example, considering that neural networks can consist of up to a billion weights, Monte Carlo simulations become inefficient to solve such problem. Hence, it is necessary to develop analytical methods which give accurate estimations of model properties.

In earlier work \cite{kryzhanovsky2021analytical}, we developed the analytical $m$-vicinity method which gives good estimations of critical temperature for 3d lattices with long-range interactions. In our definition, the $m$-vicinity is a set of spin configurations with a given value of the magnetization $m$. The central limit theorem allows us to estimate the density of states in the $m$-vicinity by its Gaussian approximation. This estimation was demonstrated to be effective for 3d disordered systems \cite{bogoslovskiy2024phase, bogoslovskiy2025disordered}. In \cite{kryzhanovsky2021analytical}, we derived analytical expressions within the Gaussian approximation for homogeneous models. Here, the $m$-vicinity method is extended to the more general case of inhomogeneous models with ferromagnetic interactions. Also, we considered the special case when a model consists of two interpenetrating sublattices, for which the expressions for the compositional dependence of the critical temperature were derived. This model can be useful for simulations of bicomponent materials.

The paper is arranged as follows. The main expressions of the $m$-vicinity method are given in Section \ref{sec2}. In Section \ref{sec3}, the critical behaviour is studied within the Gaussian estimation. The model consisting of two sublattices is considered in Section \ref{sec4}. The analysis of the critical temperature estimation by the $m$-vicinity method for inhomogeneous models by comparison with MC simulations is given in Section \ref{sec5}. The results are discussed in Section \ref{sec6}.

\section{Basic expressions of $m$-vicinity method}\label{sec2}
We consider a system consisting of $N$ spins ${s_i} = \left\{ { \pm 1} \right\}, i = \overline {1,N}$. The Hamiltonian of this system is given by the following expression:
\begin{equation}
\label{hamiltonian}
{E_H} = E - mH, \quad E =  - \frac{1}{{2N}}\sum\limits_{i,j = 1}^N {{J_{ij}}{s_i}{s_j}},  \quad m = \frac{1}{N}\sum\limits_{i = 1}^N {{s_i}},
\end{equation}
where $m$ is the magnetization, $H$ is the external field and ${\bf{J}} = \left\{ {{J_{ij}} \geq 0 \mid i,j = \overline {1,N} } \right\}$ is the connectivity matrix. The matrix ${\bf{J}}$ is symmetrical and all of its diagonal elements are equal to zero. The partition function $Z$ for the above-defined Ising model can be written as follows
\begin{equation}
\label{partition_func}
Z = \sum\limits_{E,m} {D(E,m)\exp \left[ { - N K \left( {E - mH} \right)} \right]},
\end{equation}
where $K  = 1/{k_B}T$ (${k_B}$ is the Boltzmann constant) is the inverse temperature. $D(E,m)$ is the density of states, i.e., the number of configurations with given values of the energy $E$ and the magnetization $m$.

In general, the exact expressions for $D(E,m)$ are unknown. However, the energy $E$ can be considered as the sum of  $N(N - 1)/2$ weakly correlated random variables. According to the central limit theorem --- the central part of  the $D(E,m)$ function has a good agreement with the Gaussian distribution. The moments of this distribution, the average energy ${E_m}$ and its variance ${N^{ - 1}}\sigma _m^2$, can be calculated exactly (see \cite{kryzhanovsky2015generalized}):
\begin{gather}
{E_m} = {E_0}\frac{{{{\left( {N - 2n} \right)}^2} - N}}{{N(N - 1)}}, \quad n = N\frac{{1 - m}}{2},\nonumber\\
\sigma _m^2 = \frac{{16n(N - n)}}{{N(N - 1)(N - 2)(N - 3)}}\left[ {A\left( {\sigma _0^2 - \frac{{2E_0^2}}{{N - 1}}} \right) + \frac{B}{2}\sigma _\lambda ^2} \right],\nonumber\\
A = (n - 1)(N - n - 1), \quad B = {{\left( {N - 2n} \right)}^2} - N + 2 \label{moments}.
\end{gather}
In exps. (\ref{moments}), ${E_0}$ is the ground state energy, $\sigma _0^2$ is the energy variance over all configurational space, and $\sigma _\lambda ^2$ is the variance of the local field acting on a spin in the ground state (for the homogeneous lattice $\sigma _\lambda ^2 = 0$). These quantities can be found from the elements of the matrix ${\bf{J}}$ given by following expressions:
\begin{gather}
		{E_0} =  - \frac{1}{{2N}}\sum\limits_{i,j = 1}^N {{J_{ij}}} , \quad \sigma _0^2 = \frac{1}{{2N}}\sum\limits_{i,j = 1}^N {J_{ij}^2} ,\nonumber\\	
		\sigma _\lambda ^2 = \frac{1}{{2N}}{\sum\limits_{i = 1}^N {\left( {\sum\limits_{j = 1}^N {{J_{ij}}} } \right)} ^2} - \frac{1}{{2{N^2}}}{\left( {\sum\limits_{i,j = 1}^N {{J_{ij}}} } \right)^2}.
\label{matrix parameters}
\end{gather}
In the limit of infinite lattice ($N \to \infty $), the exps. (\ref{moments}) take a much simpler form:
\begin{equation}
\label{moments2}
{E_m} = {E_0}{m^2}, \quad \sigma _m^2 = \sigma _0^2{(1 - {m^2})^2} + 2\sigma _\lambda ^2{m^2}(1 - {m^2}). 
\end{equation}

The main idea of the $m$-vicinity method is to estimate $D(E,m)$ using the following Gaussian distribution:
\begin{equation}
\label{gauss}
D(E,m)=\frac{\sqrt{N}}{\sqrt{2\pi}\sigma_m}C^n_N \text{ exp} \left[-\frac{1}{2}N\left(\frac{E-E_m}{\sigma_m}\right)^2 \right],
\end{equation}
where ${E_m}$ and $\sigma_m^2$ are given by exps. (\ref{moments2}). Substituting (\ref{gauss}) in  (\ref{partition_func}), replacing the sum with an integral and using the Stirling's formula, we obtain up to insignificant constant:
\begin{equation}
\label{integral}
	Z \sim \int\limits_0^1 {dm\int\limits_{ - \infty }^\infty  {dE} \;{e^{ - N{\kern 1pt} f(m,E)}}} ,
\end{equation}	
where
\begin{gather}
\label{free energy norm}
		f(m,E) = S(m) + K(E - mH) + \frac{1}{2}{\left( {\frac{{E - {E_m}}}{{{\sigma _m}}}} \right)^2}, \\	
		S(m) = \frac{{1 + m}}{2}\ln \left( {\frac{{1 + m}}{2}} \right) + \frac{{1 - m}}{2}\ln \left( {\frac{{1 - m}}{2}} \right).
\end{gather}

	The integral (\ref{integral}) can be calculated by the saddle-point method. The equations $\partial f(m,E)/\partial m = 0$ and $\partial f(m,E)/\partial E = 0$ yield the values of the energy $U=E$ and the magnetization $M=m$ in the saddle point. Since this method gives the value of integral up to the term $O(N^{-1})$, its results can be considered exact in the limit of the infinite lattice $N\to \infty$. Applying the saddle point method,  the free energy of the system can be expressed:	
\begin{equation}	
\label{free energy}
F = \ln{Z} = N f(M,U) + O(\ln{N^{-1}}) \approx \left(N>>1\right) \approx N f(M,U).
\end{equation}

From exp.(\ref{free energy}), it follows that the function $f(m,E)$, in the limit of the infinite lattice, is in fact the free energy per one spin. Therefore, by maximizing $f(m,E)$,  the equilibrium values of the magnetization $m$ and the energy $E$ at a given temperature $K$ can be obtained. First of the extremum conditions $\partial f(m,E)/\partial E = 0$ gives the relation:
\begin{equation}	
\label{energy relation}
E = E_0m^2-K\sigma^2_m.
\end{equation}
Applying the second extremum condition $\partial f(m,E)/\partial m = 0$ and substituting (\ref{energy relation}) in the obtained relation, the equation of state reads
\begin{equation}
\label{equation of state}
\frac{1}{2}\ln \left( {\frac{{1 + m}}{{1 - m}}} \right) = 2mK \left[ {\left| {{E_0}} \right| - K \sigma _0^2\left( {1 - {m^2}} \right) + K \sigma _\lambda ^2\left( {1 - 2{m^2}} \right)} \right] + K H.
\end{equation}
The equation (\ref{equation of state}) indirectly sets the temperature dependence of the magnetization $m(K)$. When $\sigma_0=0$ and $\sigma_\lambda=0$, the equation (\ref{equation of state}) reduces to the well-known result from the mean field theory \cite{baxter2016exactly}.

\section{Critical properties}\label{sec3}
In this Section, the critical properties of the Ising model within the framework of the $m$-vicinity are studied.

\paragraph*{Critical temperature}
Considering the case of zero external field $H = 0$, the equation (\ref{equation of state}) can be rewritten in a more convenient form:
\begin{equation}
	\label{equation of state2}
		\frac{1}{{2mq}}\ln \left( {\frac{{1 + m}}{{1 - m}}} \right) = b\left[ {1 - b\theta(m)} \right], \quad \theta(m) = 1 - {m^2} - p(1 - 2{m^2}),
\end{equation}		
where three dimensionless quantities were introduced
\begin{equation}
\label{dimensionless}
		b = K \frac{{\sigma _0^2}}{{\left| {{E_0}} \right|}}, \quad q = \frac{{2E_0^2}}{{\sigma _0^2}}, \quad p = \frac{{\sigma _\lambda ^2}}{{\sigma _0^2}}.
\end{equation}
The quantity $q$ can be considered as the “effective number of neighbors”. In the case of $d$-dimensional lattice with only nearest-neighbor isotropic interactions, $q = 2d$. When the lattice is homogeneous and all non-zero elements of the matrix ${\bf{J}}$ are equal, the quantity q is equal to the number of connections per spin. When the lattice is inhomogeneous and/or has anisotropic interactions, the quantity $q$ can take the fractional value below the number of connections per spin. The quantity $p$ is the reduced variance of the local field in the ground state.

At the critical temperature, when $m \to 0$ ($b  \to {b _c}$),  the equation (\ref{equation of state2}) takes a form
\begin{equation}
\label{crit_eq}
		\frac{1}{q} = b_c\left[ {1 - b_c\left( {1 - p} \right)} \right].
\end{equation}
The equation (\ref{crit_eq}) has no real solution for the variable $b_c$, when $q < 4\left( {1 - p} \right)$. This implies that the $m$-vicinity method predicts no phase transition in the model when this condition holds true. Solving the eq. (\ref{crit_eq}), the expression is obtained for the critical temperature:
\begin{equation}
\label{crit_temp}	
K_c=b_c\frac{\left| E_0 \right|}{\sigma^2_0}, \quad \text{where} \quad {b_c} = \frac{1}{{2(1 - p)}}\left( {1 - \sqrt {1 - \frac{{4(1 - p)}}{q}} } \right).
\end{equation}

The accuracy of the exp. (\ref{crit_temp}) in the case $p=0$ was studied in \cite{kryzhanovsky2021analytical}. For three-dimensional lattices with long-range interactions, the relative error of the critical temperature estimation is below 5\% and the exp. (\ref{crit_temp}) was found to be significantly more accurate than the mean-field predictions. For lattices on a hypercube with nearest-neighbor interactions, the accuracy improves with increasing lattice dimension. The case $p\ne0$ is studied in Section 4 of the current paper.

\paragraph*{Temperature dependencies of physical quantities}
When $K  < {K _c}$ ($m = 0$), temperature dependencies of free energy $f$, internal energy $U=\partial f/\partial K$ and the specific heat $C=-K^2 ({d^2}f/d{K^2})$ are obtained from the exps. (\ref{free energy norm}) and (\ref{energy relation}): 
\begin{equation}
\label{temp dep}
		f =  - \ln 2 - \frac{1}{2}{K ^2}\sigma _0^2 , \quad U =  - K \sigma _0^2 ,\quad C = {K^2}\sigma _0^2.
\end{equation}
	
When $K  \ge {K _c}$ ($m \ne 0$), the temperature dependence of the magnetization is set by the equation \ref{equation of state2}, which can be resolved for the variable $b$:
\begin{equation}
\label{temp}
		b = \frac{1}{{2\theta(m)}}\;\left[ {1 - \sqrt {1 - \frac{{2\theta(m)}}{{qm}}\;\ln \left( {\frac{{1 + m}}{{1 - m}}} \right)\;} \;} \right]. 
\end{equation}
The expressions for temperature dependencies of physical quantities take a form
\begin{gather}
		f = S(m) - \frac{1}{2}qb{m^2} - \frac{1}{4}q{b^2}{\left( {1 - {m^2}} \right)^2} - \frac{1}{2}qp{b^2}{m^2}(1 - {m^2}), \label{free energy2} \\	
		U = {E_0}\left[ {{m^2} + b{{\left( {1 - {m^2}} \right)}^2} + 2bp{m^2}\left( {1 - {m^2}} \right)} \right], \\	
		\sigma _E^2 = \sigma _m^2 + 4E_0^2\frac{{{m^2}\left( {1 - {m^2}} \right){{\left[ {1 - 2b\left( {1 - {m^2}} \right) + 2pb(1 - 2{m^2})} \right]}^2}}}{{1 - qb\left( {1 - {m^2}} \right)\left[ {1 - b\left( {1 - 3{m^2} - p + 6p{m^2}} \right)} \right]}}  \label{specific heat},	
\end{gather}
where $\sigma _E^2 =  - {d^2}f/d{K ^2}$ is the variance of the energy related to the specific heat by the relation $C = {K ^2}\sigma _E^2$ and the exp. (\ref{specific heat}) obtained by differentiating the eq. (\ref{equation of state2}) with respect to the variable $K$. Setting the values for the parameter $m$ from 0 to 1 and calculating the corresponding values of $b$ from the exp. (\ref{temp}), one can obtain the temperature dependencies of $f$, $U$ and $C$ from exp. (\ref{free energy2})-(\ref{specific heat}), when $K > K_c$.

From (\ref{temp dep}-\ref{specific heat}) it follows that the free and internal energies are continuous at $K  = {K _c}$, and the specific heat experiences a jump-like discontinuity. Indeed, expression (\ref{specific heat}) is valid at $K > {K _c}$, from which in the limit $m \to 0$( $K  \to {K _c}$) follows the maximum value of the specific heat:
\begin{equation}
	\sigma^2_E=\sigma^2_0+6E^2_0\frac{\left(1-2b_c+2pb_c\right)^2}{1-3qb^2_c(1-2p)},
\end{equation}
where ${b_c}$ is defined in (\ref{crit_temp}). Comparing with (\ref{specific heat}), one can see that there is a jump in the energy variance, which corresponds to a jump in the specific heat:
\begin{equation}
\label{specific heat jump}
	\Delta C = \frac{3}{2}q^2b^2_c\frac{\left(1-2b_c+2pb_c\right)^2}{1-3qb^2_c(1-2p)}.
\end{equation}
Note that, when $q = 4(1 - p)$, the specific heat jump is zero $\Delta C=0$. 

The exp. (\ref{specific heat jump}) has a good agreement with Monte Carlo simulation results for the hypercubic lattices with nearest neighbor interactions (see Table \ref{specific heat table}). It is well-known that three-dimensional lattices have an infinite discontinuity of the specific heat at the critical temperature \cite{kryzhanovsky2023modeling, haggkvist2007ising}. Hence, in this case, the $m$-vicinity method has a qualitative disagreement with actual system behavior. However, we found that the dependencies of the maximum value of the specific heat on the parameter $q$ has the same curve shape as the results of Monte-Carlo simulation for the lattices of the finite size (see \cite{kryzhanovsky2021analytical}).

\begin{table*}
\caption{Comparison of estimations obtained by the $m$-vicinity method and results of Monte Carlo simulations for the jump of the specific heat $\Delta C$ at the critical temperature $K_c$.  The results are presented for hypercubic lattices with nearest neighbor interactions.\label{specific heat table}}
\begin{tabular*}{\textwidth}{@{\extracolsep\fill}lllll@{}}\toprule
&\multicolumn{2}{@{}c@{}}{\textbf{Jump of the specific heat at the critical temperature, $\Delta C$}} \\
\cmidrule{2-3}
\textbf{Lattice dimension, $d$} & \textbf{Simulation results}  & \textbf{The prediction of the $m$-vicinty method calculated by the formula (\ref{specific heat jump})}   \\
\midrule
4 & 2.4554\cite{lundow2023revising}  & 2.1213  \\
5 & 1.8703\cite{lundow2015discontinuity}  & 1.8469    \\
6 & 1.7403\cite{lundow2015discontinuity}  & 1.7394    \\
7 & 1.686\cite{lundow2015discontinuity}    & 1.6824    \\
\bottomrule
\end{tabular*}
\end{table*}

\paragraph*{Critical exponents}
To obtain the critical exponents let us introduce a variable
\begin{equation}
	t = \frac{K-K_c}{K_c}.
\end{equation}
Then, the spontaneous magnetization that arises near the critical point ($t \to 0$, $K  > {K _c}$ ) is described by the expression following from (\ref{equation of state2}):
\begin{equation}	
\label{beta1}
		m \approx {A_1}\sqrt t , \quad  {A_1} = {\left( {\frac{{qb_c^2\left( {1 - p} \right) - 1}}{{qb_c^2\left( {1 - 2p} \right) - \frac{1}{3}}}} \right)^{\frac{1}{2}}}
\end{equation}			
This expression differs greatly from the dependence $M\sim t^{1/8}$ in the two-dimensional Ising model ($q = 4$), but is qualitatively consistent with the expression $M = 2\sqrt t $ from the van der Waals theory and the expression $M = \sqrt {3\,t} $ from the mean field theory, to which it passes in the limit $q >  > 1$ and $p = 0$.  From exp. (\ref{beta1}) follows the value of the critical exponent $\beta=1/2$. 

When the condition $q = 4(1 - p)$ is met, the critical behavior of spontaneous magnetization changes and is described by the expression:
\begin{equation}
\label{beta2}
		m \approx {A_2}t, \quad {A_2} = {\left( {\frac{1}{{2{b_c}\left( {1 - 2p} \right) - \frac{1}{3}}}} \right)^{\frac{1}{2}}}.
\end{equation}
Hence, the critical exponent $\beta=1$, when $q = 4(1 - p)$.
	
Because the specific heat exhibits a finite jump at the critical temperature, we cannot define the critical exponent $\alpha$ in a classical way. The alternative approach to define $\alpha$ should be used \cite{baxter2016exactly}
\begin{equation}
	f_+(K)-f_-(K) \sim t^{2-\alpha}, \text{ when } t \to 0,
\end{equation}
where $f_+(K)$ and $f_-(K)$ are free energies $K<K_c$ and $K>K_c$, respectively, analytically extended beyond their domains. From (\ref{temp dep}) and (\ref{free energy2}) it follows that
\begin{equation}
	f_+(K)-f_-(K) \sim m^4, \text{ when } t \to 0.
\end{equation}
Since the critical behavior of the magnetization follows the relations (\ref{beta1}) and (\ref{beta2})	, the critical exponent takes a classical value $\alpha=0$, when $q \ne 4(1 - p)$ and changes to $\alpha=-2$, when $q = 4(1 - p)$.
	
Differentiating equation (\ref{equation of state}), we obtain the temperature dependence for the susceptibility $\chi=\partial m/ \partial H$:
\begin{equation}
\label{chi}
		\chi  = \frac{{K \left( {1 - {m^2}} \right)}}{{1 - qb\left( {1 - {m^2}} \right)\left[ {1 - b\left( {1 - 3{m^2} - p + 6p{m^2}} \right)} \right]}}.	
\end{equation}		
From (\ref{chi}) it follows that the susceptibility $\chi$, when passing through the critical point, has an infinite discontinuity described by the expression:
\begin{equation}
		{\chi ^{ - 1}} \approx \left\{ {\begin{array}{*{20}{c}}
\begin{array}{l}
 - 2\left| {{E_0}} \right|\,t\;\sqrt {1 - 4(1 - p)/q}  ,\quad t \le 0\\

\end{array}\\
{\quad 4\left| {{E_0}} \right|t\;\sqrt {1 - 4(1 - p)/q},\quad t \ge 0}
\end{array}} \right.\quad 
\end{equation}	
The difference from the analogous expression of the mean field theory is the presence of the factor $\sqrt {1 - 4(1 - p)/q} $, which becomes 1 at $q >  > 1$. Note that this behavior of the susceptibility is consistent with the Curie-Weiss law, and it follows the classical value of the critical exponent $\gamma=1$.

In the case of $q = 4(1 - p)$, the susceptibility behavior changes:
\begin{equation}
		{\chi ^{ - 1}} \approx \left\{ {\begin{array}{*{20}{c}}
\begin{array}{l}
\left| {{E_0}} \right|\,{t^2}\; ,\quad t \le 0\\

\end{array}\\
{\quad  - 2\left| {{E_0}} \right|{t^2}\; ,\quad t \ge 0}
\end{array}} \right.\quad.	
\end{equation}
Therefore, the critical exponent $\gamma = 2$, when $q \ne 4(1 - p)$.

It is easy to see that the scaling hypothesis is confirmed by our method. Indeed, expanding in (\ref{equation of state}) in terms of small parameters $m$ and $t$, we obtain the relation $H = H(m,\;t)$, which can be represented in classical form ${\beta _c}H = m{\left| m \right|^{\delta  - 1}}{h_s}(t{m^{ - 2}})$ with a critical exponent $\delta  = 3$ and a scaling function of the form:
\begin{equation}
{h_s}(x) = \frac{1}{3} - qb_c^2\left( {1 - 2p} \right)\, - x\,\left[ {1 - qb_c^2\left( {1 - p} \right)} \right].
\end{equation}

When $q > 4(1 - p)$, the $m$-vicinity method gives the classical set of critical exponents $\alpha = 0$, $\beta = 1/2$, $\gamma = 1$ and $\delta = 3$. When $q = 4(1 - p)$, the jump of the specific heat vanishes and the critical behaviour changes: $\alpha = -2$, $\beta = 1$, $\gamma = 2$ and $\delta = 3$. Nevertheless, scaling relations $\gamma=\beta(\delta-1)$ and $\alpha+2\beta+\gamma=2$ hold true in both cases.

\section{Model of bicomponent material}\label{sec4}

In this Section, we consider the model consisting of two interacting sublattices. This case of an inhomogeneous lattice is of particular interest because of application for describing bicomponent composite materials. Let us assume that to describe the energy interactions in the pure first material, one can use an Ising-like lattice model with the interaction matrix ${\bf{W}}$, and to describe the pure state of the second material - a model with the matrix ${\bf{Y}}$. We denote the energy of the ground state and the effective number of neighbors of the matrix ${\bf{W}}$ as ${E_{01}}$ and ${q_1}$, respectively, and the corresponding parameters for the matrix ${\bf{Y}}$ as ${E_{02}}$ and ${q_2}$. Since both materials in their pure form are homogeneous, the variance of the local field in the ground state for both matrices is zero, that is ${p_1} = {p_2} = 0$.

We consider a lattice model of a composite material consisting of two spin sublattices. If both interacting spins belong to the same sublattice, then their interaction is determined by the corresponding elements of the matrices ${\bf{W}}$ or ${\bf{Y}}$. If the spins belong to different sublattices, then their energy interaction is determined as the geometric mean of the corresponding elements of the connectivity matrices, i.e. ${J_{ij}} = \sqrt {{W_{ij}}{Y_{ij}}} $. This definition is reasonable for most potentials of intermolecular bonds (Coulomb, dipole-dipole, Lennard-Jones, etc.). Denoting by $x \in \left[ {0;1} \right]$ the fraction of spins belonging to the first sublattice, the number of spins in the first sublattice is $w = Nx$, and in the second sublattice is $N - w$. Let us introduce the numbers ${R_i}$, $i \in 1,2...N$, where ${R_i} = 1$, if the spin $i$ belongs to the first sublattice, and ${R_i} = 0$, if the spin belongs to the second sublattice. Using this notation, the connectivity matrix elements $J_{ij}$ can be represented as follows:
\begin{equation}
\label{element}
		{J_{ij}} = {W_{ij}}{R_i}{R_j} + \sqrt {{W_{ij}}{Y_{ij}}} \left[ {{R_i}\left( {1 - {R_j}} \right) + {R_j}\left( {1 - {R_i}} \right)} \right] + {Y_{ij}}\left( {1 - {R_i}} \right)\left( {1 - {R_j}} \right).	
\end{equation}
	
To obtain the dependence of the model properties on the composition of the material, it is necessary to find expressions for the dependences of the connectivity matrix parameters on the fraction of the first sublattice $x$, i.e. for the functions ${E_0}(x)$, $q(x)$ and $p(x)$. Since the distribution of spins over the sublattices is random, the parameter values are not clearly defined as functions of $x$. Therefore, we can find only the average values of these functions $\left\langle {{E_0}(x)} \right\rangle $, $\left\langle {q(x)} \right\rangle $ and $\left\langle {p(x)} \right\rangle $. However, for large values of $N$, the deviations from these average values will tend to zero. Therefore, in further calculations, we omit the sign of the average $\left\langle {} \right\rangle $, considering these functions to be deterministic. 

The dependence of the average values of the connection matrix parameters of the composite ${E_0}(x)$, $q(x)$ and $p(x)$ in the limit $N \to \infty $ is given by the expressions (the derivation is given in the Appendix \ref{app1}):
\begin{gather}
{E_0}(x) = {E_{01}}{x^2} - 2{G_{11}}x\left( {1 - x} \right) + {E_{02}}{\left( {1 - x} \right)^2}, \quad q(x) = \frac{{2E_0^2(x)}}{{\sigma _0^2(x)}}, \nonumber \\
\sigma _0^2(x) = \frac{{2E_{01}^2}}{{{q_1}}}{x^2} + 2{G_{22}}x\left( {1 - x} \right) + \frac{{2E_{02}^2}}{{{q_2}}}{\left( {1 - x} \right)^2}, \quad p(x)=\frac{{\lambda (x) - 2E_0^2(x)}}{{\sigma _0^2\left( x \right)}}, \nonumber \\
\lambda (x) - \sigma _0^2\left( x \right) = {G_{2200}}{x^3} + 2{G_{1210}}{x^2}\left( {1 - x} \right) + {G_{1111}}x\left( {1 - x} \right) + 2{G_{1012}}x{\left( {1 - x} \right)^2} + {G_{0022}}{\left( {1 - x} \right)^3}, \nonumber \\
{G_{rm}} = \frac{1}{{2N}}\sum\limits_{i,j} {W_{ij}^{r/2}Y_{ij}^{m/2}}, \quad {G_{rmab}} = \frac{1}{{2N}}\sum\limits_{i,j,l} {W_{ij}^{r/2}W_{il}^{m/2}Y_{ij}^{a/2}Y_{il}^{b/2}{\chi _{jl}}} \label{corr},
\end{gather}
where $\chi _{jl}$ is equal to 1, when $j=l$, otherwise taking the value of 0. 

In the general case, one needs to calculate the quantities of $G$, which represent correlations between interactions in two sublattices. In the special case, when the elements of the sublattices connection matrices are proportional to each other ${Y_{ij}} = \mu {W_{ij}}$ (effectively, $q_1=q_2$), expressions (\ref{corr}) are significantly simplified
\begin{gather}
{E_0}(x) = {E_{01}}{\left[ {x + \sqrt \mu  \left( {1 - x} \right)} \right]^2}, \quad q(x) = {q_1}{\Omega ^2}(x), \nonumber \\
p(x) = 1 + \left( {{q_1} - 1} \right)\Omega (x) - {q_1}{\Omega ^2}(x), \quad \Omega (x) = \frac{{{{\left[ {x + \sqrt \mu  \left( {1 - x} \right)} \right]}^2}}}{{x + \mu \left( {1 - x} \right)}} \label{comp1}.
\end{gather}
Substituting (\ref{comp1}) in (\ref{temp}), we obtain the expression for compositional dependence of the critical temperature:
\begin{equation}
\label{proportional}
K_c(x)=\left[ \left|E_{01}\right| \left[x+\sqrt{\mu}\left(1-x\right) \right]^2 \left(1+\sqrt{\frac{4(q_1-1)}{q_1\Omega(x)}-3} \right) \right]^{-1}
\end{equation}

In the case $\mu  = 0$, i.e. when the second sublattice represents vacancies (voids), the expressions for the compositional dependencies of the parameters of the connectivity matrices take an even simpler form:
\begin{equation}
	{E_0}\left( x \right) = {E_{01}}{x^2}, \quad q\left( x \right) = {q_1}{x^2}, \quad p\left( x \right) = \left( {1 - x} \right)\left( {{q_1}x + 1} \right).
\end{equation}	
Then, the compositional dependence of the critical temperature is given by the following expression
\begin{equation}
\label{voids}
K_c(x)=	\left[ \left|E_{01}\right| x^2 \left(1+\sqrt{\frac{4(q_1-1)}{q_1 x}-3} \right) \right]^{-1}.
\end{equation}

\section{Monte-Carlo simulation}\label{sec5}
In order to estimate the accuracy of $m$-vicinity method and to establish its limitations, we carried out MC simulations for inhomogeneous three-dimensional lattices with nearest neighbor and next-to-nearest neighbor interactions (the case of homogeneous lattices was previously studied in \cite{kryzhanovsky2021analytical}).  Simulated lattices consisted of two sublattices, just like in the bicomponent model described in Section \ref{sec4}. Let us denote the values of the connectivity matrix elements determining the nearest neighbors interactions as $W_1$ and $Y_1$, and the values of the elements  determining the next-to-nearest neighbor interactions as $W_2$ and $Y_2$. All other elements of connectivity matrices are set to zero. Because we consider the cubic lattice, each spin has 6 nearest neighbors and 12 next-to-nearest neighbors.
 
According to the above-written description, the lattices were generated with sizes 20x20x20 and 32x32x32 with periodic boundary conditions, where the compositional parameter $x$ was varied from 0 to 1 with a step $0.1$. The parameters $\left|E_0\right|$, $q$ and $p$ of the connectivity matrices $\textbf{J}$ for the generated lattices with the size $N=32^3$ are shown in the Table \ref{parameters table}. The parameters for lattices with the size $N=20^3$ have less than 1\% deviation from the table values.

\begin{table*}
\caption{The parameters of the connectivity matrices of simulated latttices. \label{parameters table}}
\begin{tabular*}{\textwidth}{@{\extracolsep\fill}lllllll@{}}\toprule
&\multicolumn{3}{@{}c@{}}{\textbf{$W_1=1$, $W_2=1$, $Y_1=0$, $Y_2=0$}} 
&\multicolumn{3}{@{}c@{}}{\textbf{$W_1=1$, $W_2=0.5$, $Y_1=0.4$, $Y_2=0.2$}} \\
\cmidrule{2-4}\cmidrule{5-7}
$x$ & $\left|E_0\right|$  & $q$ & $p$ & $\left|E_0\right|$ & $q$ & $p$  \\
\midrule
0 & -  & - & - & 2.4 & 16 & 0  \\
0.1 & -  & - & - & 2.69 & 15.16 & 0.44  \\
0.2 & -  & - & - & 2.99 & 14.7 & 0.68  \\
0.3 & -  & - & - & 3.31 & 14.47 & 0.79  \\
0.4 & 1.44  & 2.87 & 4.91 & 3.65 & 14.41 & 0.82  \\
0.5 & 2.25  & 4.50 & 4.99 & 4.00 & 14.49 & 0.78 \\
0.6 & 3.24  & 6.47 & 4.73 & 4.37 & 14.67 & 0.69 \\
0.7 & 4.41  & 8.82 & 4.08 & 4.75 & 14.91 & 0.57 \\
0.8 & 5.76  & 11.52 & 3.08 & 5.15 & 15.23 & 0.40 \\
0.9 & 7.29  & 14.58 & 1.72 & 5.57 & 15.59 & 0.22 \\
1.0 & 9.00  & 18.00 & 0.00 & 6.00 & 16.00 & 0.00 \\
\midrule
&\multicolumn{3}{@{}c@{}}{\textbf{$W_1=1$, $W_2=1$, $Y_1=0.5$, $Y_2=0.5$}} 
&\multicolumn{3}{@{}c@{}}{\textbf{$W_1=2$, $W_2=0$, $Y_1=1$, $Y_2=0.5$}} \\
\cmidrule{2-4}\cmidrule{5-7}
$x$ & $\left|E_0\right|$  & $q$ & $p$ & $\left|E_0\right|$ & $q$ & $p$  \\
\midrule
0 & 4.5  & 18.00 & 0.00 & 6.00 & 16.00 & 0.00  \\
0.1 & 4.88  & 17.50 & 0.26 & 5.68 & 13.33 & 0.11  \\
0.2 & 5.28  & 17.19 & 0.42 & 5.44 & 11.20 & 0.12  \\
0.3 & 5.69  & 17.02 & 0.51 & 5.26 & 9.54 & 0.10  \\
0.4 & 6.11  & 16.96 & 0.54 & 5.16 & 8.28 & 0.07  \\
0.5 & 6.56  & 16.99 & 0.53 & 5.12 & 7.36 & 0.05 \\
0.6 & 7.01  & 17.09 & 0.47 & 5.16 & 6.72 & 0.05 \\
0.7 & 7.49  & 17.25 & 0.39 & 5.26 & 5.26 & 0.06 \\
0.8 & 7.98  & 17.45 & 0.29 & 5.44 & 6.05 & 0.06 \\
0.9 & 8.48  & 17.71 & 0.15 & 5.68 & 5.96 & 0.04 \\
1.0 & 9.00  & 18.00 & 0.00 & 6.00 & 6.00 & 0.00 \\
\bottomrule
\end{tabular*}
\end{table*}

MC simulation was carried out in the vicinity of the critical temperature $K_c$ in the absence of the external field. The inverse temperature $K$ was varied with a step $10^{-4}$. The spins were updated in their typewrite order by the standard Metropolis criteria. For each temperature the first $10^4N$ MC steps were completed without accumulation of statistics, in order to achieve the equilibrium state. The total number of MC steps at a given temperature was $10^5$ per spin. The energetic and magnetic quantities ($E$, $E^2$, $m^2$, $m^4$) were measured every $N$ MC steps.

\begin{figure} 
\centering 
\includegraphics[width=0.5\textwidth]{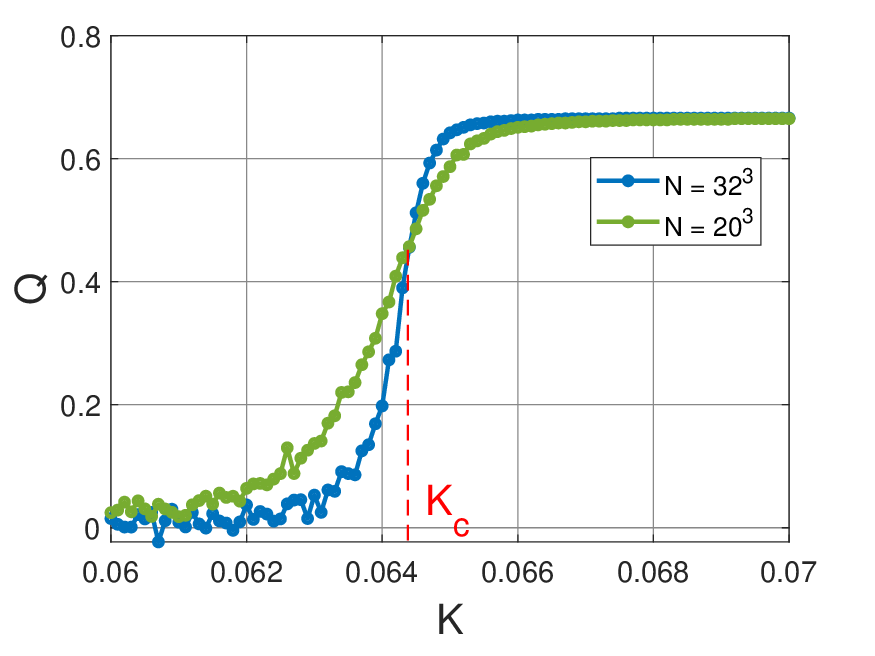}
\caption{Temperature dependencies of the Binder cumulant for lattices of different sizes. The lattices have interaction parameters $W_3=1$, $W_4=1$ ($x = 1$). The intersection point of these curves yields the critical temperature for the infinite lattice ($N\to \infty$). } 
\label{fig1} 
\end{figure}

\begin{figure*} [!t]
\centering
    \begin{subfigure}[b]{0.48\textwidth}
        \centering
        \includegraphics[width=\textwidth]{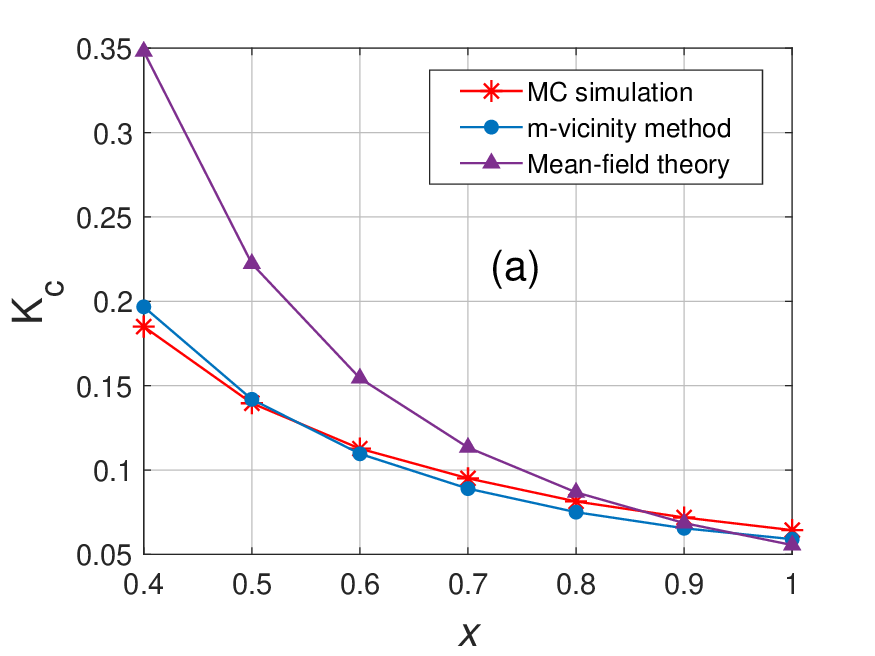}
    \end{subfigure}
    \hfill 
    \begin{subfigure}[b]{0.48\textwidth}
        \centering
        \includegraphics[width=\textwidth]{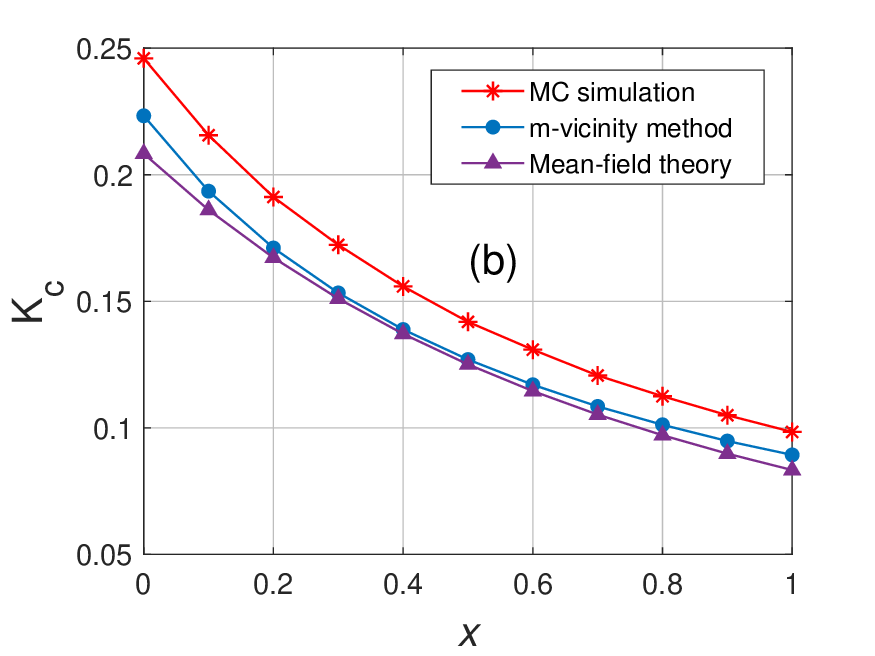}
    \end{subfigure}
    \begin{subfigure}[b]{0.48\textwidth}
        \centering
        \includegraphics[width=\textwidth]{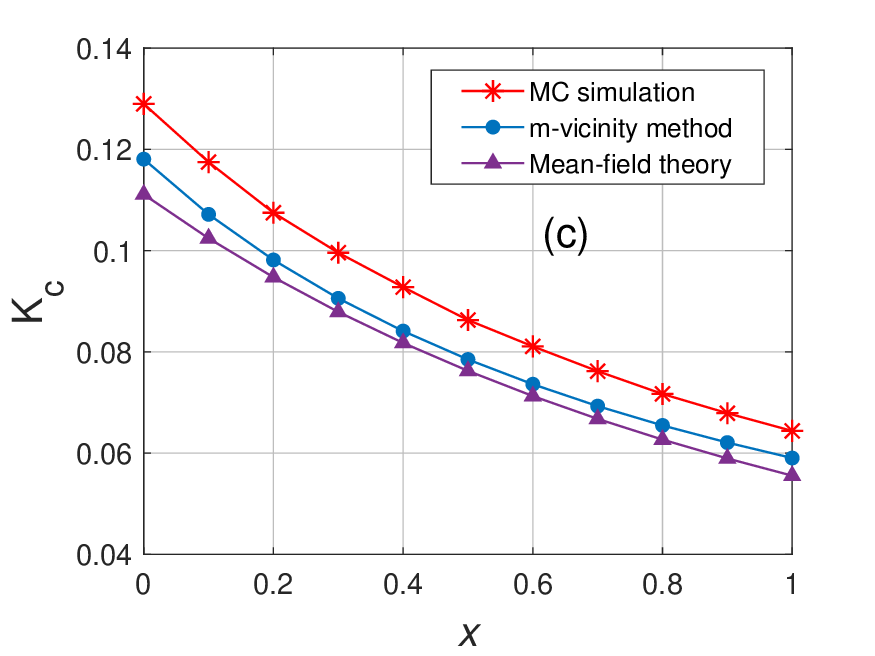}
    \end{subfigure}
    \hfill 
    \begin{subfigure}[b]{0.48\textwidth}
        \centering
        \includegraphics[width=\textwidth]{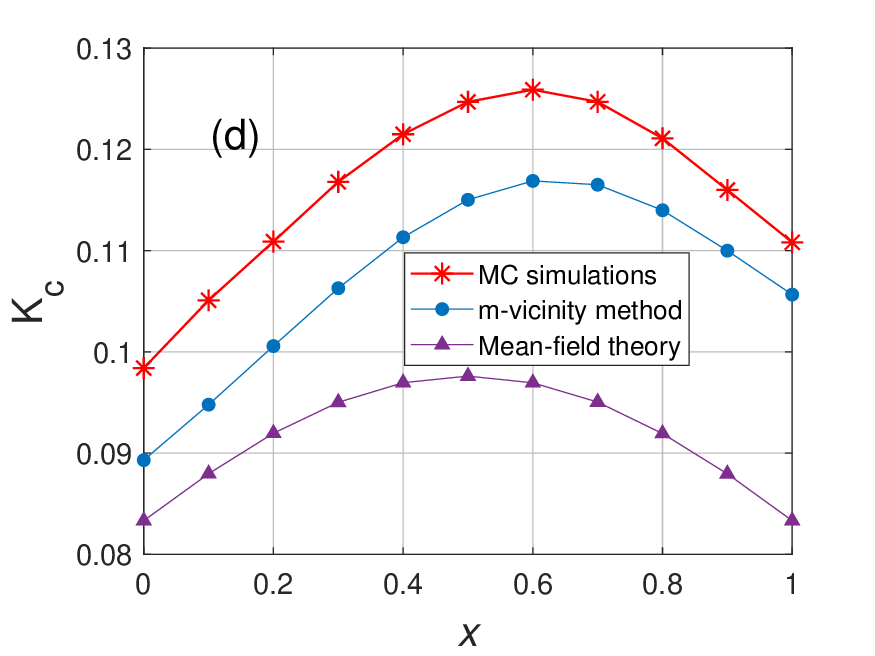}
    \end{subfigure}
    \caption{Compositional dependencies of the critical temperature $K_c$. Comparison of MC simulation, $m$-vicinity method, and mean-field theory. The parameters of the sublattices are (a) $W_1=1$, $W_2=1$, $Y_1=0$, $Y_2=0$; (b) $W_1=1$, $W_2=0.5$, $Y_1=0.4$, $Y_2=0.2$; (c) $W_1=1$, $W_2=1$, $Y_1=0.5$, $Y_2=0.5$; (d) $W_1=2$, $W_2=0$, $Y_1=1$, $Y_2=0.5$.}
\label{fig2}    
\end{figure*}

To determine the critical temperature $K_c$ for the infinite size lattice ($N\to\infty$), we used Binder cumulants \cite{binder1981finite}. The Binder cumulant $Q$ is defined by the following expression:
\begin{equation}
Q=1-\frac{\left\langle {m^4} \right\rangle}{3 {\left\langle {m^2} \right\rangle}^2}.
\end{equation}
By plotting the temperature dependencies of Binder cumulants $Q(K)$ for lattices with different sizes, the critical temperature $K_c$ can be found by the intersection point of these dependencies. This method is illustrated in Fig. \ref{fig1}. By comparing with well-known values of critical temperature for square and cubic lattices with nearest neighbor interaction, we found that the absolute error of this estimation is no more than $10^{-4}$. The comparison of the critical temperature values $K_c$, obtained by MC simulation, with estimations of our $m$-vicinity method and the mean-field theory $K_c=1/\left(2\left|E_0\right|\right)$ is given in Fig. \ref{fig2}.

In Fig. \ref{fig2}(a), the compositional dependence of the critical temperature $K_c(x)$ is shown in the case, when the lattice contains vacancies, which fraction is inversly proportional to the parameter $x$, i.e., $Y_1=0$ and $Y_2=0$. We did not consider lattices with $x<0.4$, because the huge amount of vacancies leads to large fluctuations in the temperature dependencies of the physical quantities. The data marked by blue circles was calculated using the exp. (\ref{voids}) with $\left|E_{01}\right|=9$ and $q_1=18$. As one can see from Fig. \ref{fig2}(a), the accuracy of our $m$-vicinity method is significantly higher than the mean-field theory, except for one point when $x=0.9$ and $q=14.58$. Note that $K_c$ obtained by the $m$-vicinity method is larger than the exact values when $q<5$ (this is an approximate estimate). This fact coincides with results obtained by us previously for homogeneous lattices \cite{kryzhanovsky2021analytical}. 

The compositional dependences of the critical temperature $K_c(x)$, when the elements of the connection matrices for sublattices are proportional to each other ${Y_{ij}} = \mu {W_{ij}}$ ($\mu>0$), are shown in Figs. \ref{fig2}(b, c). Here, the $m$-vicinity estimation was obtained by the exp. (\ref{proportional}) with $\left|E_{01}\right|=6$, $q_1=16$, $\mu=0.4$ in Fig. \ref{fig2}(b)  and $\left|E_{01}\right|=9$, $q_1=18$, $\mu=0.5$ in Fig. \ref{fig2}(c). As one can see, the exp. (\ref{proportional}) has a better agreement with simulation results compared with predictions of the mean-field theory across the entire range of the compositional parameter $x$. Moreover, the relative error of the $m$-vicinity estimation changes only slightly with variation of $x$. Meanwhile, the mean-field theory works better near the middle range of $x$.

The most general case, when the elements of connectivity matrices of sublattices are not proportional to each other, is shown in Fig. \ref{fig2}(d). In this case, the $m$-vicinity estimation of the critical temperature $K_c$ should be determined by the exp. (\ref{crit_temp}), where compositional dependencies $E_0(x)$, $q(x)$ and $p(x)$ are given by exp. (\ref{corr}). The compositional dependence of the critical temperature $K_c(x)$ presented in Fig. \ref{fig2}(d) is non-monotonic and it has a maximum at $x=0.6$. The $m$-vicinity method correctly predicts the position of the maximum value of $K_c(x)$. Meanwhile, the mean-field theory incorrectly places the maximum at $x=0.5$.

Overall, the relative error of the $m$-vicinity method for estimation of the critical temperature $K_c$ is less than 6\% for all simulated lattices. Our expressions give significantly more accurate qualitative and quantitative estimations of $K_c(x)$ than the mean-field theory, while requiring almost the same computational effort.

\section{Conclusions}\label{sec6}
The analytical expressions were obtained for inhomogeneous Ising models within the $m$-vicinity estimation. We also derived the expressions for compositional dependence of the critical temperature in the case when the model consists of two interpenetrating sublattices. The accuracy of the critical temperature estimation was analysed by comparison with MC simulations. We found that the maximum relative error of our method was less than 6\% for 3d lattices with nearest neighbor and next-to-nearest neighbor interactions. The agreement with simulation is expected to increase for lattices with larger dimensions. One can find expression (\ref{crit_temp}) to be useful when fast estimation of the critical temperature is needed.

Our method, in most cases when $q > 4(1 - p)$, predicts the mean-field values of critical exponents, which agrees with the behaviour of the Ising model with dimensions $d\ge4$ \cite{aizenman1986critical}. However, in contrast with the mean-field theory, the $m$-vicinity method gives a good estimation for the magnitude of the specific heat jump at the critical point (see the Table \ref{specific heat table}). The interesting result is that the critical behaviour changes when $q = 4(1 - p)$. In this case, the critical exponents are similar to those observed in spin glass models \cite{klein1991series}. We expect that this critical behaviour will be more evident in our approach when negative (antiferromagnetic) interactions are present. The extension of the $m$-vicinity method for models with competing ferromagnetic and antiferromagnetic interactions will be the topic of our future work.

%\backmatter
\section*{Author contributions}

Vladislav Egorov: Conceptualization, Methodology, Software, Writing --- Original draft preparation, Investigation. Stepan Osipov: Data curation, Visualization, Writing --- Reviewing and Editing, Investigation.

\section*{Acknowledgments}
The work is supported by the State Program of Research Center Kurchatov Institute – SRISA, project no. FNEF-2024-0001.

\section*{Data Availability Statement}
The data that support the findings of this study are openly available in “figshare” at 10.6084/m9.figshare.30294409.

\bibliography{wileyNJD-AMA}

\appendix

\section{The derivation of expressions for compositional dependence of parameters of connectivity matrix\label{app1}}
\vspace*{12pt}

Here, we give the derivation for exp. (\ref{corr}). Let us number by the index $k = 1,2...C_N^w$, $w=xN$, all possible ways to distribute the spins between two sublattices for a given $x$. Then, the average value of the ground-state energy over these $C_N^w$ distributions is given by
\begin{equation}
{E_0}(x) = \frac{1}{{C_N^w}}\sum\limits_{k = 1}^{C_N^w} {E_0^{(k)}}  =- \frac{1}{{2NC_N^w}}\sum\limits_{i,j = 1}^N {\sum\limits_{k = 1}^{C_N^w} {J_{ij}^{(k)}} } .
\end{equation}
 Substituting (\ref{element}) into this expression and summing over indices $i$ and $j$, we obtain:
\begin{equation}
\label{A2}
{E_0}(x) = \frac{1}{{C_N^w}}\left( {{E_{01}}\sum\limits_k {R_i^{(k)}R_j^{(k)}} {\chi _{ij}} - 2{G_{11}}\sum\limits_k {R_i^{(k)}\left( {1 - R_j^{(k)}} \right)} {\chi _{ij}} + {E_{02}}\sum\limits_k {\left( {1 - R_i^{(k)}} \right)\left( {1 - R_j^{(k)}} \right){\chi _{ij}}} } \right),
\end{equation}
where we use the notation ${G_{rl}} = \frac{1}{{2N}}\sum\limits_{i,j} {W_i^{r/2}Y_j^{l/2}}$ and take into account the symmetry with respect to the permutation of indices $i$ and $j$. Here, $\chi _{ij}=1$, when $i \ne j$, and $\chi _{ij}=0$, when $i = j$. Note that the sums over $k$ do not depend on the specific values of $i$ and $j$, provided that $i \ne j$ (${\chi _{ij}} = 1$), since the spin distribution over sublattices is uniform and random. Since the diagonal elements of matrices ${\bf{W}}$ and ${\bf{Y}}$ are equal to zero, this allows us to perform the summation over $i$ and $j$ independently of $k$, replacing the sums of the matrix elements with the quantities ${E_{01}}$, ${E_{02}}$, and ${G_{11}}$.

	Sums over $k$ can be easily calculated. Indeed, if a term in the sum contains $r$ factors of the form ${R^{(k)}}$ and $z$ factors of $1 - {R^{(k)}}$, then it is nonzero and equal to one if specific $r$ spins belong to the first sublattice and other specific $z$ spins belong to the second sublattice. The total number of configurations satisfying this condition for a given $x$ is equal to the number of ways to select $w - r$ spins from $N - r - z$. As a result, we obtain:
\begin{equation}
\label{A3}	
	\sum\limits_k {R_{{i_1}}^{(k)}R_{{i_2}}^{(k)}...R_{{i_r}}^{(k)}\left( {1 - R_{{j_1}}^{(k)}} \right)\left( {1 - R_{{j_2}}^{(k)}} \right)...\left( {1 - R_{{j_z}}^{(k)}} \right)} {\chi _{{i_1}{i_2}...{i_r}{j_1}{j_2}...{j_z}}} = C_{N - r - z}^{w - r}.
\end{equation}	

Substituting (\ref{A3}), the expression (\ref{A2}) takes a form
\begin{equation}	
\label{A4}
{E_0}(x) = {E_{01}}\frac{{w\left( {w - 1} \right)}}{{N\left( {N - 1} \right)}} - 2{G_{11}}\frac{{w\left( {N - w} \right)}}{{N\left( {N - 1} \right)}} + {E_{02}}\frac{{\left( {N - w} \right)\left( {N - w - 1} \right)}}{{N\left( {N - 1} \right)}}.
\end{equation}
Replacing $w = Nx$, in the limit $N \to \infty$, we finally obtain
\begin{equation}
\label{A5}	
{E_0}(x) = {E_{01}}{x^2} - 2{G_{11}}x\left( {1 - x} \right) + {E_{02}}{\left( {1 - x} \right)^2}.
\end{equation}

Similar to the calculations above, we derived an expression for the sum of the squares of the matrix elements $\sigma _0^2(x) = \frac{1}{{2N}}\sum\limits_{i,j} {J_{ij}^2} $ 
\begin{equation}
	\sigma _0^2(x) = \frac{{2E_{01}^2}}{{{q_1}}}\frac{{w\left( {w - 1} \right)}}{{N\left( {N - 1} \right)}} + 2{G_{22}}\frac{{w\left( {N - w} \right)}}{{N\left( {N - 1} \right)}} + \frac{{2E_{02}^2}}{{{q_2}}}\frac{{\left( {N - w} \right)\left( {N - w - 1} \right)}}{{N\left( {N - 1} \right)}},
\end{equation}
which in the limit $N \to \infty$ takes a form
\begin{equation}
\label{A7}	
	\sigma _0^2(x) = \frac{{2E_{01}^2}}{{{q_1}}}{x^2} + 2{G_{22}}x\left( {1 - x} \right) + \frac{{2E_{02}^2}}{{{q_2}}}{\left( {1 - x} \right)^2}.
\end{equation}
The compositional dependence $q(x)$ can be obtained by definition (\ref{dimensionless}):
\begin{equation}
\label{A8}	
	q(x) = \frac{{2E_0^2(x)}}{{\sigma _0^2(x)}}.	
\end{equation}

To calculate $p(x)$, one needs to find an expression for the following quantity:
\begin{equation}
\label{A9}
\lambda \left( x \right) = \frac{1}{{2N}}{\sum\limits_i {\left( {\sum\limits_j {{J_{ij}}} } \right)} ^2} = \frac{1}{{2N}}\sum\limits_{i,j} {{J_{ij}}} \sum\limits_l {{J_{il}}} \left( {{\chi _{jl}} + {\delta _{jl}}} \right) = \sigma _0^2(x) + \frac{1}{{2N}}\sum\limits_{i,j,l} {{J_{ij}}{J_{il}}{\chi _{jl}}},  
\end{equation}
where $\delta$ is the Kronecker delta. Substituting ((\ref{element})) into (\ref{A9}) and averaging over all possible arrangements of sublattices, we obtain
\begin{equation}
\label{A10}
\frac{1}{{2N}}\left\langle {\sum\limits_{i,j,l} {{J_{ij}}{J_{il}}{\chi _{jl}}} } \right\rangle  
 = \frac{1}{{C_N^w}}\left[ {{G_{2200}}C_{N - 3}^{w - 3} + 2{G_{1210}}C_{N - 3}^{w - 2} + {G_{1111}}\left( {C_{N - 3}^{w - 2} + C_{N - 3}^{w - 1}} \right) + 2{G_{1012}}C_{N - 3}^{w - 1} + {G_{0022}}C_{N - 3}^w} \right],
 \end{equation}
where the notation is introduced	
\begin{equation}	
\label{A11}
		{G_{rzab}} = \frac{1}{{2N}}\sum\limits_{i,j,l} {W_{ij}^{r/2}W_{il}^{z/2}Y_{ij}^{a/2}Y_{il}^{b/2}{\chi _{jl}}} .
\end{equation}
In the limit $N \to \infty$, expression (\ref{A10}) takes the form		
\begin{equation}
\label{A12}
\frac{1}{{2N}}\left\langle {\sum\limits_{i,j,l} {{J_{ij}}{J_{il}}{\chi _{jl}}} } \right\rangle  = 
 {G_{2200}}{x^3} + 2{G_{1210}}{x^2}\left( {1 - x} \right) + {G_{1111}}x\left( {1 - x} \right) + 2{G_{1012}}x{\left( {1 - x} \right)^2} + {G_{0022}}{\left( {1 - x} \right)^3}.
  \end{equation}
Let us note that
\begin{equation}			
{G_{2200}} = \frac{1}{{2N}}\sum\limits_{i,j,l} {{W_{ij}}} {W_{il}}{\chi _{jl}} = 2E_{01}^2\left( {1 - \frac{1}{{{q_1}}}} \right), \quad		
		{G_{0022}} = \frac{1}{{2N}}\sum\limits_{i,j,l} {{Y_{ij}}} {Y_{il}}{\chi _{jl}} = 2E_{02}^2\left( {1 - \frac{1}{{{q_2}}}} \right).
\end{equation}
Finally, $p(x)$ can be calculated by the following expression 
\begin{equation}		
p(x) = \frac{{\lambda (x) - 2E_0^2(x)}}{{\sigma _0^2\left( x \right)}}.
\end{equation}
where $\lambda (x)$, $E_0(x)$ and $\sigma _0^2\left( x \right)$ are defined by exp. (\ref{A5}), (\ref{A7}-\ref{A9}) and (\ref{A12}).

\end{document}